\documentclass{iau}
\usepackage{graphicx}
\usepackage[T1]{fontenc}

\title[Everything we'd like to do with LSST data, but we don't know (yet) how] 
{Everything we'd like to do with LSST data, but we don't know (yet) how}

\author[Ivezi\'{c}, Connolly \& Juri\'{c}]   
{\v{Z}eljko Ivezi\'{c}$^1$, Andrew J. Connolly$^1$ \& Mario Juri\'{c}$^1$
}

\affiliation{
$^1$ Department of Astronomy, University of Washington, \\ Box 351580, Seattle, WA 98195-1580, USA\\ 
                               email: {\tt ivezic@astro.washington.edu}
}

\pubyear{2017}
\volume{325}  
\pagerange{1--10}
\jname{Astroinformatics}
\editors{M. Brescia, eds.}

\begin{document} 
\maketitle

\begin{abstract}
The Large Synoptic Survey Telescope (LSST), the next-generation optical imaging survey 
sited at Cerro Pachon in Chile, will provide an unprecedented database of astronomical
measurements. The LSST design, with an 8.4m (6.7m effective) primary mirror, a 9.6 sq. deg.
field of view, and a 3.2 Gigapixel camera, will allow about 10,000 sq. deg. of sky to be 
covered twice per night, every three to four nights on average, with typical 5-sigma depth 
for point sources of  $r$=24.5 (AB). With over 800 observations in $ugrizy$ bands over a 
10-year period, these data will enable a deep stack reaching $r$=27.5 (about 5 magnitudes 
deeper than SDSS) and faint time-domain astronomy. The measured properties of newly 
discovered and known astrometric and photometric transients will be publicly reported 
within 60 sec after observation. The vast database of about 30 trillion observations 
of 40 billion objects will be mined for the unexpected and used for precision experiments 
in astrophysics. In addition to a brief introduction to LSST, we discuss a number of 
astro-statistical challenges that need to be overcome to extract maximum information 
and science results from LSST dataset.  
\end{abstract}

\firstsection 

\section{Introduction \label{sec:lsst}}

\begin{figure}[t!]
\begin{center}
\includegraphics[width=0.99\textwidth, angle=0]{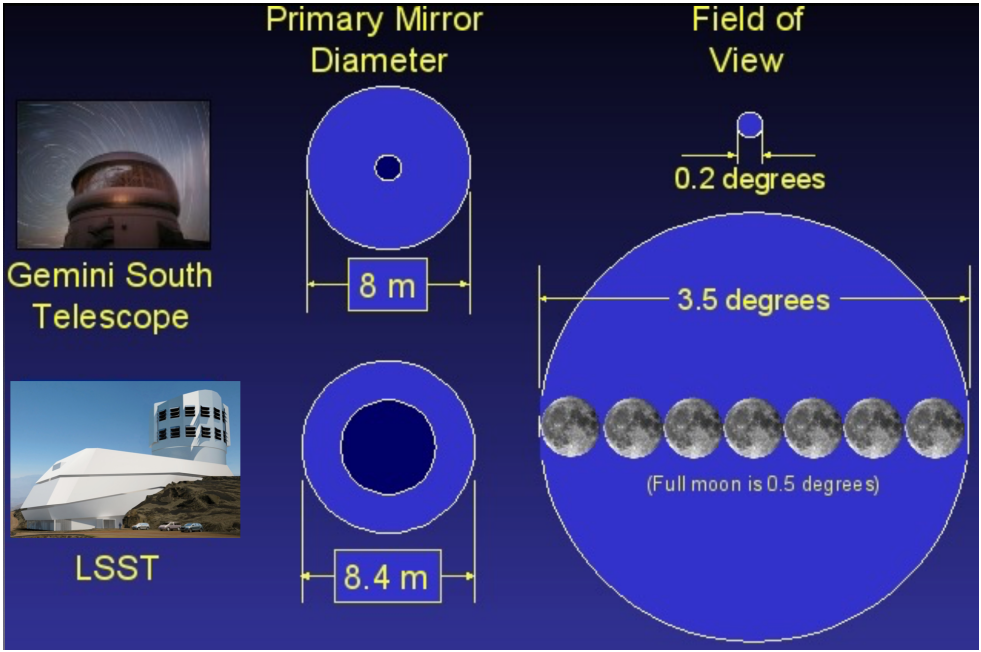} 
\vskip -0.3in
\caption{
A comparison of the primary mirror size and the field-of-view size for LSST 
and Gemini South telescopes. The product of the primary mirror size and 
the field-of-view size, the so-called \'etendue (or grasp), a characteristic 
that determines the speed at which a system can survey a given sky area 
to a given flux limit, is much larger for LSST. Figure courtesy of Chuck Claver.}
\label{fig:Gemini}
\end{center}
\end{figure}

The last decade has seen fascinating observational progress in optical imaging surveys.
The SDSS dataset is currently being greatly extended by the ongoing surveys such as 
Pan-STARRS (Kaiser et al. 2010) and the Dark Energy Survey (Flaugher 2008). The Large Synoptic 
Survey Telescope (LSST)  is the most ambitious survey currently planned in the visible band
(for a brief overview, see Ivezi\'{c} et al. 2008a). The unparalleled LSST survey power is due to 
its large \'etendue (see Figure~\ref{fig:Gemini}). 

The goals of the LSST are driven by four key science themes: probing dark energy and dark matter, 
taking an inventory of the Solar System, exploring the transient optical sky, and mapping the 
Milky Way. The LSST will be a large, wide-field ground-based system designed to obtain multiple 
images covering the sky visible from Cerro Pach\'{o}n in Northern Chile. 
The system, with an 8.4m (6.7m effective) primary mirror, a 9.6 deg$^2$ field 
of view, and a 3.2 Gigapixel camera, will allow, on average, about 10,000 deg$^2$ of sky to be covered 
using pairs of 15-second exposures in two photometric bands every three nights, 
with a typical 5$\sigma$ depth for point sources of $r\sim24.5$. The system is designed to 
yield high image quality as well as superb astrometric and photometric accuracy \footnote{For 
detailed specifications, please see the LSST Overview Paper, \cite{Ivezic08LSST}, and the LSST 
Science Requirements Document (LSST Science Collaboration 2011)}. 
The survey area will cover 30,000 deg$^2$ with $\delta<+34.5^\circ$, and will be imaged 
multiple times in six bands, $ugrizy$, covering the wavelength range 320--1050 nm. About 90\% of the 
observing time will be devoted to a deep-wide-fast survey mode which will observe an
18,000 deg$^2$ region over 800 times (summed over all six bands) during the anticipated 
10 years of operations, and yield a coadded map to $r\sim27.5$. These data will result in 
databases including about 20 billion galaxies and a similar number of stars, and will 
serve the majority of science programs. The remaining 10\% of the observing time 
will be allocated to special programs such as a Very Deep and Fast time-domain 
survey, the details of which are still being defined. More details about various science programs 
that will be enabled by LSST data can be found in the LSST Science Book (LSST Science Collaboration 2009) 
and at  the LSST website (www.lsst.org). 

\begin{figure}[t!]
\begin{center}
\includegraphics[width=0.99\textwidth, angle=0]{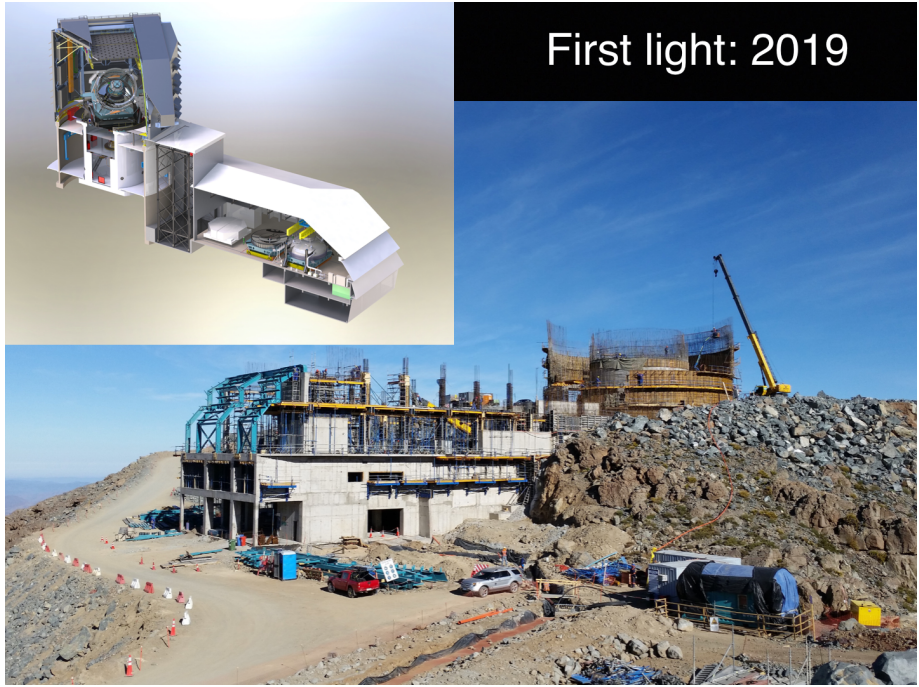} 
\vskip -0.2in
\caption{
The inset in the top left corner shows a cut-away render of the LSST Observatory building. 
The rest of the figure shows a photograph of the LSST summit at the time of this 
Symposium (September 2016). First light for LSST is expected with a 144 Mpix engineering
camera in 2019, and with the full 3.2 Gpix camera in 2020. For more 
photographs, see https://www.lsst.org/gallery/image-gallery}
\label{fig:summit2019}
\end{center}
\end{figure}

First light for LSST is expected at the end of  2019 with a small commissioning camera (144 Mpix), with 
the full 3.2 Gpix camera integrated in 2020. The construction phase of LSST, funded by the U.S. 
National Science Foundation and Department of Energy, started in 2015 and is progressing
according to the planned schedule (see Figure~\ref{fig:summit2019}). 

\section{LSST Data Analysis Challenges}

The LSST project will deliver data products that will enable a large number of cutting-edge science 
programs (Juri\'{c} et al. 2016). Nevertheless, depending on the topic, the path from LSST data 
products to science results and journal papers may sometimes require additional challenging
analysis work. These challenges, representative of the era of Big Data, stem from:
\begin{itemize}
\item Large data volumes (petabytes)
\item Large numbers of objects (billions)
\item Highly multi-dimensional spaces (thousands)
\item Unknown statistical distributions 
\item Time-series data (irregular sampling)
\item Heteroscedastic errors, truncated, censored and missing data
\item Unreliable quantities (e.g. unknown systematics and random errors)\\
\end{itemize}

{\it Everything we'd like to do with LSST data, but we don't know
  (yet) how} is a catchy title but somewhat inaccurate. First, we most
certainly do not include here ``everything'', and second, we and our
LSST colleagues already have at least some ideas for how to approach
most of problems discussed below.  We hope that this contribution will
help motivate others to join us in this thinking, and to engage in
the work needed to maximize LSST's scientific yields.\\

To begin and stimulate this conversation, we have selected a few topics where substantial
preparatory work is needed to optimally analyze datasets at the LSST scale. These are:
\begin{enumerate}
\item Interpreting spectral energy distributions (SEDs)
\item Identifying moving objects
\item Characterizing and classifying variable stars 
\item Understanding systematic measurement uncertainties
\item Characterizing astrophysical simulations and astrophysical
  systematics
\item Devising new or enhanced algorithms to process LSST data.\\
\end{enumerate}

We emphasize that many other members of LSST Science Collaborations contributed to
the formulation of this, by all means, incomplete list. In the remainder of this section, 
we discuss these topics in a bit more detail.

\subsection{Interpretation of spectral energy distributions (SEDs)}

Efficient and robust interpretation of time-resolved multi-band photometry for ``billions and 
billions'' of objects is bound to yield unprecedented science results. A combination of 
required measurement precision and relatively wide bandpasses will require careful
interpretation of LSST data. 

A broad-band photometric system, such as LSST, aims to deliver calibrated in-band flux
\begin{equation}
\label{Fb}
             F_b = \int{F_\nu(\lambda) \phi_b(\lambda) d\lambda},
\end{equation}
where $F_\nu(\lambda)$ is specific flux of an object {\it at the top} of 
the atmosphere and $\phi_b(\lambda)$ is the normalized system response 
for the given band,
\begin{equation}
\label{PhiDef}
\phi_b(\lambda) = {\lambda^{-1} S_b(\lambda) \over \int{\lambda^{-1} S_b(\lambda) d\lambda}}
\end{equation}
(the $\lambda^{-1}$  term reflects the fact that CCDs are photon-counting devices). 
Here, $S_b(\lambda)$ is the overall atmosphere + system throughput
\begin{equation}
\label{SDef}
         S_b(\lambda) = S_b^{sys}(\lambda) \times S_b^{atm}(\lambda). 
\end{equation}

Numerous science programs can be cast as constraining the possible forms of the true SED 
$F_\nu(\lambda)$ given the measured broad-band fluxes, $F_b$, and the 
normalized system response, $\phi_b(\lambda)$, with
$b=(u,g,r,i,z,y)$. Because of the integration 
over broad bandpasses, forward modeling using a trial SED (either empirical or model based) is 
typically superior to ``correcting data'' (fluxes, positions, sizes). Examples of such programs, 
where SEDs presumably depend on relevant astrophysical parameters, include
\begin{enumerate}
\item photo-z algorithms: the observed galaxy and quasar SEDs depend on the redshift of an intrinsic SED
(due to expansion of the universe, source evolution, and intergalactic extinction; see e.g. Bolzonella et al. 2000);
\item photometric parallax for stars, where measured colors can be used to constrain the effective
          temperature  and luminosity (e.g. Juri\'{c} et al. 2008);
\item photometric metallicity for stars (trained using spectroscopic metallicities, see Ivezi\'{c} et al. 2008b); and
\item interstellar extinction along the line of sight for stars in the Milky Way disk (see, e.g., Berry et al. 2012).\\
\end{enumerate}

There are a number of open issues that are being worked on by the community: 
\begin{itemize}
\item What are the relative advantages and disadvantages of machine
  learning methods compared to methods based on fitting
          SED templates (both empirical and simulated). How can we
          incorporate ancillary data (and priors) within the photo-z
          methods, for example utilizing angular cross-correlation of 
         photometric and spectroscopic samples of galaxies (Newman 2008)? 
\item What are the impacts of heteroscedastic noise, priors, and truncated and censored data? 
\item How much will ``per-visit processing'' of LSST data help (due to varying bandpasses $\phi_b(\lambda)$ 
          because of the unavoidable variations in $S_b^{atm}(\lambda)$)? 
\item What is the best way to handle posterior probability density functions (pdfs), how much is gained compared 
         to simple (e.g. maximum likelihood) point estimates, what are the optimal compression algorithms for pdfs, etc.?  
\item How should the parameter covariances be handled (the same question is also valid pretty much everywhere 
          else below)? 
\end{itemize}

\subsection{Moving objects}

The catalogs generated by LSST will increase the known number of small
bodies in the Solar System by a factor of 10-100, among all
populations (Jones et al. 2016). The median number of observations for
Main Belt asteroids will be on the order of 200-300, allowing sparse
lightcurve inversion to determine rotation periods, spin axes, and
shape information. The current strawman for the LSST survey strategy
is to obtain two visits of the same field per night (each ``visit''
being a pair of back-to-back 15s exposures), separated by about 30
minutes, and covering the entire observable sky every 3-4 days
throughout the observing season.

The main reason for two observations per night is to help association of observations of the same moving object from 
different nights, as follows. The typical distance between two nearby asteroids on the Ecliptic, at the faint fluxes 
probed by LSST, is a few arcminutes (counts are dominated by Main Belt asteroids). Typical asteroid motion during 
several days is much larger (of the order a degree or more) and thus, without additional information, detections of individual 
objects are ''scrambled''. However, with two detections per night, the motion vector can be estimated. The motion 
vector makes the linking problem much easier because positions from one night can be approximately extrapolated 
to future (or past) nights.

There are several interesting open questions regarding moving objects: 
\begin{itemize}
\item Cadence optimization: are two visits per night really needed?
  Would perhaps a substantial increase in the computing power solve
  the association problem with just a single detection per night?
\item How robust and efficient would be a full Bayesian approach for
  characterizing the orbits of asteroids (see, e.g., Virtanen et al. 2001)? 
\item How computationally hard is it to deploy shift-and-coadd method for KBOs and
  more distant objects on LSST-scale dataset?
\item What are the most robust and efficient methods for sparse lightcurve inversion of several million asteroids? 
\end{itemize}

\subsection{Variable stars}

Early in the survey, LSST will be discovering about 100,000 variable
stars per night at high Galactic latitudes (Ridgway et al. 2014), and
probably many more at low latitudes (but the forecast is less
certain). The total number of variable stars to be discovered by LSST
is of the order several hundred million (the total number of detected and measured
stars will be about 20 billion). In addition, about 1000 new
supernovae are expected to be discovered every observing night. A
number of statistical questions need to be answered for the full
exploitation of this dataset:
\begin{itemize}
\item How to best distinguish regular (periodic) from irregular
  variability when the data are sampled irregularly and when the
  variability may be wavelength dependent? 
\item How to distinguish short from long variability timescales? 
\item What are the best methods for the robust detection of
  variability, and for anomaly detection\footnote{Extensive tools for 
doing both template-matched and ``model-independent'' detection 
of variability have been recently developed in the context of LIGO.}?  
Recent developments in compressed sensing and deep learning have the potential to
  revolutionize the analysis of variability and transient
  detection. By exploiting the sparseness of the data and careful choice of the models
  that might be fit to these data, it maybe possible to characterize
  and classify sources in a way that is both flexible and robust to
  noise.
\item What are the best methods for characterization and
  classification of a broad range of variability (especially in case
  of sparse data early in the survey)? How do machine learning methods
  compare to light curve template-based methods?  Are there metrics that will
  enable a general classification scheme for identifying sources that
  might need follow up observations?
\item Is it possible to further optimize the cadence to enhance discoveries
   and characterization of variable stars? 
\item What is the impact of heteroscedastic noise, astrophysical
  priors, and truncated and censored data?
\item Can light curve and objects characterization and classification be done directly in database?  
\end{itemize}


\subsection{Systematic measurement uncertainties}

Due to the large number of objects in LSST samples, many science programs, including 
cosmology, will be sensitive to systematic errors. In many cases the volume of the
available data will mean that systematics are the dominant source of uncertainty (that is,
given billions of objects measured a thousand times, how do we know that sqrt(N) will still 
work in this regime?). The primary goals include:
\begin{enumerate}
\item ensuring that the astrometry can be measured with statistical and systematic errors at the miliarcsec level,
\item ensuring that the photometry can be measured with statistical and systematic errors at the milimag level,
\item measuring galaxy shapes (e.g., for use in cosmic shear analysis) with the PSF known across
  the focal plane to a level where the autocorrelation of PSF residuals is smaller than $10^{-7}$.\\
\end{enumerate}

These effects will need to be quantified as functions of position on the sky, position on
the focal plane, observing conditions (e.g., atmospheric seeing, sky brightness), and object 
properties (e.g., brightness, colors, size). Some of the open questions include:
\begin{itemize}
\item How can the impact of unknown SEDs be quantified? 
\item What is the impact of the atmosphere (due to variable seeing and transmissivity, 
          differential chromatic refraction and intrinsic stochasticity)? 
\item How can we robustly quantify both multiplicative and additive errors in galaxy shear measurements? 
\item How can we control systematic errors in photometric redshifts? 
\end{itemize}

\subsection{Astrophysical simulations and astrophysical systematics}

The expected precision of the LSST measurements and their resulting
constraints on cosmological and astrophysical models requires the
development of simulation and modelling tools of equal or better
precision. These tools will need to provide predictions for what the LSST
will observe (in order to define effective survey strategies for the
LSST),  interpret the observations in the context of physically
motivated models, and generate multiple realizations of a simulation
to characterize the uncertainties or covariances within the
cosmological and astrophysical constraints.

Simulations come in a number of different flavors: instrument
simulations that model the optical performance of the LSST (e.g., the
impact of the atmosphere, telescope, and camera on the image quality),
cosmological simulations that model the evolution in structure within
the universe, and mock or synthetic catalogs which provide
realizations of the sky with appropriate distributions of the
astrophysical properties of sources and their uncertainties.

The scale of the required simulations range from approximations of the
technical parameters that describe the LSST (OpSim; Delgado et
al. 2014), through single realizations of an image from an LSST sensor (DESC 2015)
and large scale mock catalogs of asteroids (Jones et al.\
2016), to full simulations of representative volumes of LSST data
(DESC 2015). Many of the required simulation tools are in
development (e.g.\ Connolly et al. 2014). There remain, however, a
number of challenges before these tools become widely accepted by the 
community.

\begin{itemize}
\item How do we support the generation of large scale simulations? The
  computational resources required to generate cosmological
  simulations, and in particular series of simulations for
  characterizing the covariance of cosmological models, are large and
  could exceed the resources available to individual investigators.
\item How do we share simulations in a manner similar to the
  availability of observation data? Often the sizes of useful simulated data
  sets must exceed those of the observational datasets and are already
  approaching the PB scale. Transferring the generated simulations or
  mock catalogs from supercomputing centers to where they might be
  analyzed will stress academic network capacities.
\item What is the impact of baryonic effects on dark matter halo
  profiles? The current generation of hydrodynamical simulations do
  not simulate large cosmological volumes. Approximations, where lower
  resolution or dark matter only models are used to identify regions
  of interest in the simulation that are then re-simulated at higher
  resolution, can lead to biases in any derived correlations as they
  are not representative volumes of the universe.
\item What are the main feedback mechanisms in galaxy formation and
  what is the best way to handle nonlinear galaxy bias?
\item How can we best address intrinsic alignments of galaxy shapes with the density field? 
\item How can we best extract the information about the evolution of the Milky Way galaxy using 
         LSST measurements of 20 billion stars? 
\item How can we best extract the information about the evolution of the Solar System using LSST 
         measurements of a few million asteroids?  
\end{itemize}

\subsection{LSST System Enhancements and New Algorithms}

LSST is an automated {\em facility} that will deliver not only raw images, 
but also fully reduced data products (calibrated single-epoch images,
multiple flavors of co-adds, and a variety of catalogs). Its cadence will 
be optimized to enable a balanced science return across the four key science
themes (Section~\ref{sec:lsst}). Its data products have been designed to enable the
derivation of a large fraction of those results without the need for end users to fully
understand the details of the LSST instrument and data reduction, algorithms, or to begin
from raw pixel data.

To make this possible, the LSST project is making a major investment in
computing infrastructure, software, and algorithm development.  Yet it is
quite clear that more and better are always possible; even marginal
improvement in performance (of both hardware and software) could yield
significant additional science returns.  While {\em some} of the open issues
listed below are already being addressed by groups both within and outside
the LSST construction project, substantial further research could be done.

Again, these are simply the most obvious examples; the list is by no means complete.

\begin{itemize}
\item Observing strategy (cadence) optimization can yield improvements
  in total open-shutter time for the survey, but also can improve the
  utility of angular and temporal sampling functions and dithering
  patterns (Delgado et al. 2014). It is, therefore, important to
  develop a scheduling algorithm that can efficiently address
  potential evolution of the LSST observing system and evolution of
  its science drivers. The LSST Project is developing a scheduling algorithm
  that meets the survey requirements, but the complexity of the problem and
  the potential return on investment\footnote{For example, just 1\% effective improvement
  in LSST scheduling is roughly equivalent to $\sim$\$4M in operational cost.}
  argues for further research.
\item LSST does not plan to deliver specialized crowded field reductions
  or catalogs; images of crowded regions of the Milky Way will be
  processed with the same code utilized elsewhere, though perhaps with
  different priors used in object detection and deblending stages (i.e., to
  a very good approximation, every object observed towards the Galactic
  center is a star). A purpose-built (multi-epoch capable) crowded field
  code capable of dealing with LSST source densities and data volumes
  would tremendously enhance the scientific return of LSST's Galactic dataset.
\item No LSST data products have been explicitly designed to enable the
  detection and characterization of diffuse (e.g., ISM) or extremely
  low-surface brightness structures (e.g., LSB galaxies recently discovered
  by projects such as Dragonfly). Developing specialized codes to enable
such processing may add significant value to the LSST dataset.
\item Complex galaxy models (e.g., tidal tails of merged galaxies) will
  not be fit by the standard LSST pipelines. Such a tool
  would greatly help in understanding gravitational potential around
  judiciously selected galaxies.
\item Forward modeling of images on per visit basis (termed Multifit
  in LSST Data Management context) is superior to analysis of co-added
  images (because of varying observing conditions) and will be done by LSST. 
  A particularly interesting problem is one of simultaneous forward modelling of data
  from {\em different} datasets (e.g., LSST and WFIRST). While there is
  substantial ongoing development (e.g. the Tractor code, see Lang et
  al. 2016), including within LSST Project,
  many statistical and other issues\footnote{For example, blended
  objects present major algorithmic challenges and a discussion of 
  their treatment, which is currently an open research area, would warrant 
  a paper on its own.} remain open and will require
  substantial further research to find the optimal approach.
\item At the required precision level, the LSST point spread function
  (PSF) will depend on time, instrument state, source position, and
  source color (more precisely, on in-band SED shape); see Meyers \&
  Burchat (2015). Robust and precise determination of the PSF will
  therefore be a rather non-trivial undertaking. The LSST project is
  required to characterize the PSF to the degree described in the LSST 
  Science Requirements Document,
  but further improvements may be possible.
\item A shift-and-stack algorithm (for co-adding images along arbitrary
  space-time trajectories), that could be efficiently deployed for
  large datasets would likely have a major impact on outer Solar
  System science. LSST Data Management (DM) system will not deliver
  shift-and-stack pipelines or data products, but these could be easily
  built on top of the open source code LSST DM will deliver.
\item Image differencing will be used to detect transient sources in the
  LSST data stream. In order to control the false positive rate,
  new sophisticated algorithms will have to be developed to account
  for varying observing conditions (e.g., the treatment of differential
  chromatic refraction effects due to varying airmass, as well as
  color-dependent PSFs). At the same time, they will need to be fast
  enough to meet the requirement of delivering alerts within 60 seconds. The LSST project
  is developing these to enable its code deliverables, but broader
  research in this area would always be welcome.
\item The SEDs and other properties of newly discovered transients will be poorly known initially.
It is not clear yet what characterization and classification algorithms would be the best for separating 
the most interesting transients that require prompt followup from the background of much more numerous
transients which can be analyzed on much longer timescales without significant loss of science outcome.
\item While there are well developed methods for the classification of light curves of variable stars
(e.g. Richards et al. 2011), transient classification with sparse data is a much harder problem. 
\item Jointly processing data from LSST and other surveys (e.g., Euclid or WFIRST) would
certainly result in a superior dataset than the one produced individually
by either of these projects (for details see, e.g., Jain et al. 2015). It is not clear, however,
how exactly to implement these ideas in practice, especially given that the survey overlap will be 
significantly truncated (either by position on the sky, e.g., for WFIRST, or by brightness, e.g. for Gaia).
\item Finally, LSST data processing will be performed in the context of a
relatively traditional HPC-like computing facility utilizing proven, low-risk,
technologies (e.g., HT Condor, Pegasus). Similarly, LSST catalog data will be served
to the users by way of relational databases (albeit of advanced, distributed, kind). Research into 
alternative models of processing (e.g., making use of the public cloud or workflow 
systems like Apache Spark) or data storage and serving (e.g., no-SQL
databases, or next-generation experiments such as SciSQL) would be of great interest.
If successful, these efforts could significantly enhance the ability of the community
to perform affordable large-scale catalog computations or even image reprocessings.\\
\end{itemize}

A number of use cases above may be possible for the users to
run at the LSST Data Access Centers. LSST has reserved approximately 10\% of
its total capacity to enable end-user analyses and generation of added value
data products.

Furthermore, many of the use cases would be best tackled by enhancing
existing LSST pipelines or building completely new functionality on top of
the one already provided by LSST. All of the source code for the LSST
pipelines will be publically available, enabling these kinds of endeavors.

Finally, LSST Operations have been built with the assumption that, in
addition to the work within the facility, the community will make new
discoveries and breakthroughs in areas of algorithms and data products.
Such enhancements, developed by the community, can be incorporated into
standard LSST processing, thus becoming a part of the official LSST alert
streams and/or data releases.

\section{Discussion}

Due to the size and complexity of the LSST dataset, and the susceptibility of many
of its major science programs to systematics in both measured quantities
and astrophysical predictions, substantial preparatory work is
required to enable the full exploitation of the LSST dataset. The bottleneck for science
will not be the size of the dataset but instead our ability to extract
useful and reliable information from the data.

Here we have summarized some of the most obvious research directions
required to enhance the LSST science outcome. The main anticipated work
areas include:
\begin{itemize}
\item advanced astronomical digital image processing,
\item statistical modeling and analysis,
\item data mining and machine learning,
\item high performance computing,
\item astrophysical simulations, and
\item multi-dimensional and temporal data visualization. \\
\end{itemize}

LSST data analysis and the development of the fields of astro-informatics \& astro-statistics will be closely
intertwined. This synergy will open numerous opportunities for people with ``Big Data'' skills.
Prospective LSST science users, across all disciplines, should collaborate and
coordinate. By working jointly we can make the LSST great, 
and maximize the tremendous potential and science return of its dataset.

\vskip 0.1in 
{\it Acknowledgements}  

This material is based upon work supported in part by the National Science Foundation through
Cooperative Agreement 1258333 managed by the Association of Universities for Research in Astronomy
(AURA), and the Department of Energy under Contract No. DE-AC02-76SF00515 with the SLAC National
Accelerator Laboratory. Additional LSST funding comes from private donations, grants to universities,
and in-kind support from LSSTC Institutional Members. We thank Gregory Dubois-Felsmann for
his careful reading and excellent comments.


\vskip -0.1in
\begin{thebibliography}{}
\bibitem[()]{} Berry, M., Ivezi\'c, \v Z., Sesar, B., et al.~2012, Astrophysical Journal, 757, 166

\bibitem[()]{} Bolzonella, M., Miralles, J.~M. \& Pell\'{o}, R. 2000, Astronomy \& Astrophysics, 363, 476

\bibitem[()]{} Connolly, A.J., Angeli, G.Z., Chandrasekharan, S., et al. 2014, Proceedings of the SPIE, 
          Volume 9150, id. 915014

\bibitem[()]{} Delgado, F., Saha, A., Chandrasekharan, S., et al. 2014, Proceedings of the SPIE, Volume 9150, 
          id. 915015

\bibitem[()]{} DESC; Dark Energy Science Collaboration (DESC) Science
  Roadmap, 2015, http://lsst-desc.org/sites/default/files/DESC\_SRM\_V1.pdf


\bibitem[()]{} Eyer, L., Evans, D.W., Mowlavi, N., et al. 2015, \textit{{\rm ArXiv:}1502.03830}

\bibitem[Flaugher (2008)]{Flaugher08}
{Flaugher, B. 2008}, In \textit{A Decade of Dark Energy: Spring Symposium, Proceedings of
  the conferences held May 5-8, 2008 in Baltimore, Maryland. (USA). Ed. by N. Pirzkal \& H. Ferguson.}

\bibitem[Ivezi{\'c} \etal\ (2008a)]{Ivezic08LSST} {Ivezi{\'c}, {\v Z}., Tyson, J.A., Acosta, E., et al. 2008a}, \textit{{\rm ArXiv:}0805.2366}

\bibitem[()]{} Ivezi\'c, \v Z., Sesar, B., Juri\'{c}, M., et al.~2008b, Astrophysical Journal, 684, 287

\bibitem[()]{} Jain, B., Spergel, D., Bean, R., et al. 2015, \textit{{\rm ArXiv:}1501.07897}

\bibitem[()]{} Jones, R.L., Juri\'{c}, M. \& Ivezi\'c, \v Z. 2016, Proceedings of the IAU, 318, 282, \textit{{\rm ArXiv:}1511.03199}

\bibitem[()]{} Juri\'{c}, M., Ivezi\'c, \v Z., Brooks, A., et al.~2008, Astrophysical Journal, 673, 864

\bibitem[()]{} Juri\'{c}, M., Kantor, J., Lim, K-T., et al. 2016, ASP Conf Ser. in press, \textit{{\rm ArXiv:}1512.07914}

\bibitem[Kaiser \etal\ (2010)]{Kaiser2010} {Kaiser, N., Burgett, W., Chambers, K., et al. 2010}, 
\textit{Proc. SPIE 7733, Ground-based and Airborne Telescopes III}, vol. 7733, 77330E

\bibitem[()]{} Lang D., Hogg, D. W. \& Mykytyn D. 2016, Astrophysics Source Code Library (ascl:1604.008)

\bibitem[SciBook]{SciBook} {LSST Science Collaboration, 2009}, LSST Science Book, arXiv:0912.0201

\bibitem[SRD]{SRD} {LSST Science Collaboration, 2011}, LSST Science Requirements Document, http://ls.st/srd

\bibitem[()]{} Meyers, J.E. \& Burchat, P.R. 2015, Astrophysical Journal, 807, 182 

\bibitem[()]{} Newman, J.A. 2008, Astrophysical Journal, 684, 88

\bibitem[()]{} Reyes, R., Mandelbaum, R., Seljak, U., et al. 2010, Nature, 464, 256 

\bibitem[()]{} Richards, J.W., Starr, D.L., Butler, N.R., et al. 2011, Astrophysical Journal, 733, 10 

\bibitem[()]{} Ridgway, S.T., Matheson, T., Mighell, K.J., Olsen, K.A. \& Howell, S.B. 2014, Astrophysical Journal, 796, 53

\bibitem[()]{} Rockosi, C.M., Odenkirchen, M., Grebel, E.K., et al. 2002, Astronomical Journal, 124, 349

\bibitem[()]{} Virtanen, J., Muinonen, K. \& Bowell, E. 2001, Icarus, 154, 412

\bibitem[()]{} Willman, B., Dalcanton, J.J. \& Martinez-Delgado, D. 2005, Astrophysical Journal, 626, 85

\end{thebibliography}
\end{document}